\documentclass{article}
\usepackage[paperwidth=612pt,paperheight=792pt,top=72pt,right=65pt,bottom=72pt,left=65pt]{geometry}
\usepackage{multicol}
\usepackage{graphicx}
\usepackage{graphics}
\usepackage{amssymb}
\usepackage{amsmath}
\usepackage{color,graphicx}
\usepackage{color}
\usepackage{hyperref}
\usepackage{makeidx}
\hypersetup{colorlinks=true,citecolor=blue,urlcolor=blue,filecolor=magenta,bookmarks=true}
\makeatletter
	{\par\setlength{\parindent}{#3}
	\setlength{\leftmargin}{#1}       \setlength{\rightmargin}{#2}%
	\advance\linewidth -\leftmargin       \advance\linewidth -\rightmargin%
	\advance\@totalleftmargin\leftmargin  \@setpar{{\@@par}}%
	\parshape 1\@totalleftmargin \linewidth\ignorespaces}{\par}%
\makeatother
\makeatletter
\@ifundefined{Umathcode}{}{}
\@ifundefined{Umathchardef}{}{}
\makeatother
\begin{document}
\begin{center}
\title
\text {{\huge\bf  Observing the spin of free electrons in action}}
\end{center}
\begin{center}
\text {{\Large\bf (Stern-Gerlach experiment by free electrons)}}
\end{center}
\begin{center}
\noindent \textit{Patent:139350140003006698 ,Tuesday, September16, 2014}\footnote{\textit{\\ PCT Support Registration Office, Patent Name:Electron Intrinsic Spin Analyzer,Inventor: Hosein Majlesi}\\ Patent Link:\url{http://ip.ssaa.ir/Patent/SearchResult.aspx?DecNo=139350140003006698&RN=84973}}
\end{center}
\begin{center}
\noindent Hosein Majlesi \footnote{ \\ \url{hoseinmajlesinew@gmail.com} -\url{hosein.majlesi@stu.qom.ac.ir} -\url{hoseinmajlesimail@gmail.com}\\
\textit{[Department of Physics,University of Qom,Graduate Student]-Iran,Qom,Personal.Postcode:3716863914}}
\end{center}
\begin{center}
\noindent \textit{Independent Researcher}
\end{center}
\noindent \textit{}
\begin{abstract}
    Stern-Gerlach experiment by free electron is very important experiment because it answered some questions that remain unanswered for almost a century. Bohr and Pauli considered its objective observation as impossible while some other scientists considered such observation as possible. The experiment on free electrons has not been conducted so far because the high magnetic field gradient predicted there was thought as impossible to generate. This paper proves that it is not only possible but also observable using a high vacuum lamp which is deionized well. To obtain a high magnetic field gradient, it is not necessary to have a very strong magnetic field and it is possible to observe the phenomenon using a very sharp pointed magnet and adjusting the voltage in a certain distance from free electron beams. that objective observation requires your consideration of some technical points simultaneously.In this experiment, no electric field and no magnetic field does not change with time. \\*
 \end{abstract}
\begin{multicols}{2}
\noindent \textbf{\large\bf Introduction}\\*
 \noindent\text{     }  The concept of electron spin was first proposed by Samuel Goudsmit, George Uhlenbeck, and Wolfgang Pauli in 1920, suggesting a physical interpretation of particles, spinning around their own axis. They stated that an intrinsically angular momentum depends on any electron, quite independent of the orbit angular momentum. This intrinsic momentum is called electron spin.\cite{1},\cite{2} Spin is considered as a fundamental property of subatomic particles, which has no classical equivalent and it is considered a quantum  property. To help visualize the model, consider it as an object in space which continuously rotates around an axis. To describe electron spin, assume a magnetic moment. If  an electron exists in an external magnetic field with its permanent magnetic moment, its spin is expected to be quantized. It means that the spin magnetic moment and spin angular momentum will be restricted to certain orientations. There are only two intrinsic spin states for electrons. In general, the electron magnetic moment is expressed as$\ \ \mu =\frac{eg}{2m}S$  , in which \textit{e} represents the electron charge; \textit{g} is gyromagnetic ratio; \textit{m} is the mass of the electron, and \textit{S} is the electron spin operator. The constant term in electrons magnetic moment is called Bohr magneton constant. When electrons are placed in an inhomogeneous magnetic field, a force is exerted from the field.
 \begin{equation}
F=-\nabla U=-\nabla \left(\mu\ .{\rm B}\right){\rm \ };\mu=0.927\times {10}^{-23}amp.m^2
\label{eq.1}
\end{equation}
\text{   }  In 1921-1922, Otto Stern suggested that magnetic dipole moments of various atoms be measured through detection of the atomic beam deflection in an inhomogeneous magnetic field.\cite{3} he had significantly developed the techniques of atomic and molecular beams during two years of Einstein's assistance. This led him to conduct an experiment in collaboration with Walter Gerlach in which they steamed silver atoms in a furnace in a vacuum, passed them through an inhomogeneous magnetic field, and finally registered them on a screen. Indeed, electron spin in silver atoms is related to the effect of the electron spin in the last atomic orbit, and the effect of nucleus and other electrons are ignored. Lorentz force is largely inhibited due to use of atoms in Stern-Gerlach experiment. Bohr, Pauli, and Mott believed that given Bohr magneton coefficient compared to the gradient of the magnetic field, such a phenomenon may not be seen in reality. The impossibility of observation of free electron spin is a general principle and this experiment can never be expressed in terms of the classical approach.\cite{4},\cite{5},\cite{6} Bohr and Pauli also stated that:\cite{7}   
\noindent \textit{"It is impossible to observe the spin of the electron, separated fully from its orbital momentum, by means of experiments based on the concept of classical particle trajectories".}  
\noindent  In the Sixth Solvay Conference in 1930, Brillouin proposed a model of an inhomogeneous magnetic field where primary fields were symmetric and along the beam path. It was significantly different from the geometry of the Stern-Gerlach's apparatus.\cite{8} Bohr and Pauli explicitly opposed his model.\cite{9} They rejected it using a semi-classical approximation cited in Mott, Massley, and Klemperer studies\cite{10},\cite{11},\cite{12}. It should be noted that Lorentz force exists in the system and it is not removable; moreover, its direction differs from the spin force. Lorentz force is certainly greater than the spin force. Electrons are separated due to dipole spatial interaction of the magnetic field with wave function assigned to electrons. In other words, due to the limited width of the beam, they deviate to the left or right when influenced by Lorentz force, so the spin force causes the beams to separate. So wave function spin separation, resulting from the interaction of the dipole magnetic field, may fade owing to the circuit magnetic effect. It may not reflect the true nature of the spin. The combination of Newtonian mechanics and quantum mechanics in this case causes any arguments to be regarded with uncertainty. Given the relativistic nature of electron spin, it is imperative that formulation of quantum be appropriate and an analogy be drawn between classical and quantum nature of the electron spin so that more accurate data are obtained about the system and position of the particles. Measuring the system is aimed to obtain data on any state of the system. As already stated, Lorentz force and the dividing force of the spin are both involved in Stern-Gerlach experiment at the same time, and it is important to set them apart. Pauli holds a prejudiced opinion that electron spin can never be observed yet Bohr had a more cautious view and said:\cite{13}
\noindent  \textit{"I have sometimes thought of the problem of the realisation of the electron polarization, and after all I am quite prepared that such a polarization might be observable. argument tells ... not that the closer quantum-theoretical treatment will never give a positive effect".} 
\noindent When Bohr and Pauli suggested the impossibility of the observation, almost all scientists had acknowledged it, until Dehmelt conducted a controversial experiment in 1988 and gained an electron magnetic momentum outside atomic framework which ran contrary to what predicted by Bohr and Pauli.\cite{14},\cite{15} So far, many questions have been raised about the possibility of Stern-Gerlach experiment with electrons.\\*
\ignorespaces
\noindent\text{       } \noindent  \textit{\it\bf Is it really possible to measure electron spin?} 
\noindent\text{       } Various studies were carried out about how to remove Lorentz force, most notably Brillouin's model. A full analysis of Quantum Wave Theory reveals that it is quite possible. Since 1997, several papers have been published in this regard. Although Bohr and Pauli's arguments received many physicist's confirmation, it is important to take Bohr and Pauli's views into account. For example, in 1997 Batelaan, Gay,and Schwendiman wrote an article in which they used Pauli's argument to reject Brillouin's model and given the semi-classical approximation, they showed observing the electron spin possible.\cite{16} In 1998, Rutherford and Grobe drew on Dirac numerical equations to demonstrate that measurement of electron spin is hard, yet possible to be observed.\cite{17}  In 1999, Garraway and Stenholm used wave packet analysis in quantum mechanics to prove that Stern-Gerlach phenomenon is possible to be observed for free electrons.\cite{18} In 2001, Gallup and Batelaan demonstrated that given the semi-classical approximation (WKB) spin separation is possible.\cite{19} In 2011 Scot McGregor, Roger Bach and Batelaan proposed a solution in their essay based on Brillouin's model.\cite{20}.many theories have been raised in this case.
\noindent   \textit{    }  You will find a geometric and mathematical solution which predicts observing the spin of free electrons at a particular point in space using a very sharp magnet (this solution is based on Stern-Gerlach experiment by potassium atomic beams
) and then the experiment with electrons will be explained in detail. \\*
\noindent\text{       } \ignorespaces
\noindent\textbf{\large\bf Description And Technical Methods }\\*
 \noindent\text{     } As stated before, in order to separate the electron's spins, Lorentz force must be controlled and a proper and inhomogeneous magnetic field must be applied to separate the electrons' spins in practice. However, as an electron has a very low mass, it seems impossible for single electrons to control Lorentz force. So, according to Bohr and Pauli, a very high magnetic field gradient is required and any proposed solution should be able to explain the related classical approach. If it consider that electron's travel in a straight line along with X axis in the space before applying a magnetic field, you can find Lorentz force from the following equation:$\ F=e\ V\times B+eE$  , in which the direction of its first term can be found based on the speed of electrons flow and direction of the magnetic field applied. Moreover, as there is no magnetic monopole $\nabla .B=0$ therefore: $(\frac{\partial B}{\partial z})=-(\frac{\partial B}{\partial y})$  So, concerning Lorentz force and spin separating force, it is expected to get the separation from classical approach:
\begin{equation}
∆Z=\frac{1}{2}at^2+vt=\frac{1}{2}[\left(\frac{\mu }{m}\right).\nabla {\rm B]}t^2+vt
\label{eq.2}
\end{equation}
\noindent\text{       } In this equation, vt term is explicitly ignored. So, only this first term  \eqref{eq.2} would be considered meaningful. In equation \eqref{eq.2}, time is stated as $t=\frac{l}{v}$ in which \textit{L} the interaction distance between electrons and magnetic field is a constant value and $v$ stands for electron's speed. Of course, concerning the speed of electrons in vacuum, it is not much far-fetched to consider magnetic field gradient as the only influential factor in the above equation. So, regarding the numerical value of   $\frac{\mu }{m}={10}^7$ , only the value of two parameters of electron speed and magnetic field gradient is unknown in this equation \eqref{eq.2} which is very important to be found. By axis analysis you can find the following simplified form of the gradient: 
\begin{align}
\nabla \left(F.G\right)&=\left(F.\nabla \right)G+F\times \left(\nabla \times G\right)+\left(G.\nabla \right)F\nonumber \\
&\qquad+G\times (\nabla \times F)
\label{eq.3}
\end{align}
\noindent\text{       } Regarding this equation, the potential gradient phrase will be in the following simplified form:
\begin{align}
\nabla \left(\mu\ .{\rm B}\right)&=\left(\mu\ .\nabla \right){\rm B}+\mu\times \left(\nabla \times {\rm B}\right)+\left({\rm B}.\nabla \right)\mu \nonumber \\
&\qquad+{\rm B}\times (\nabla \times\mu)
 \label{eq.4}
\end{align}
\noindent\text{       } As $\mu $ is a constant value, its related differentials will be zero. Besides, as zero value of phrase  $\mu\times \left(\nabla \times {\rm B}\right)$ depends on phrase ${\rm (}\nabla \times {\rm B)}$. based on Maxwell's equations you will have:\begin{equation}
{\rm \ }\nabla \times B=\frac{1}{c^2}\frac{\partial E}{\partial t}=\frac{\omega E}{c^2}=\frac{2\pi fE}{c^2}
\label{eq.5}
\end{equation}
\noindent\text{       }  Therefore, if you use a alternative electric field with a term  $\frac{\partial E}{\partial t}\ $will not be zero but  $\frac{\partial E}{\partial t}=\omega E=2\pi fE$ so the spin force will be as follow: 
\begin{align}
F &=-\nabla \left(\mu {\rm .B}\right){\rm =}{\rm -}\left(\mu \left(\nabla {\rm B}\right){\rm +}\mu {\rm \times }\left(\frac{1}{c^2}\frac{\partial E}{\partial t}\right)\right) \nonumber \\
&\qquad = {\rm -}\left(\mu .\left(\nabla {\rm B}\right){\rm +}\mu {\rm \times }{\rm (}\frac{2\pi fE}{c^2})\right)
\label{eq.6}
\end{align}
\noindent\text{       } It can be expected that a term is added to spin separating force. However, as electric field is not an important factor in classical Stern-Gerlach experiment on silver atoms, so you will have $\nabla $$\times$B=0  Hence, only the term $\mu\ .(\nabla {\rm B)}$ is left in this experiment which causes the dispersion of beams. Electrons cannot be evaporated like silver atoms but they can be easily separated by applying an appropriate voltage from hot Tungsten in a high vacuum and appropriately deionizes condition and then lead them flow through parallel plates. The mass of an electron is much less than a silver atom mass, so the speed of electrons is much greater than the speed of silver atoms. Electric field is also unavoidable which will somehow complicate the issue. Hence, Lorentz force is highly influential and cannot easily be ignored in this calculations. Because much stronger electric field is required to remove Lorentz force from the system which may call for relativistic relations. That is why it's not been included in calculations so far.\\
\noindent\text{  }\ignorespaces
\textbf{ Note:}\footnote{\textit{This was not mentioned in the old version (arXiv:1504.07963v1 and arXiv:1504.07963v2)}} Equation\eqref{eq.6}Demonstrates the effect of changes in magnetic field over time. These changes, in turn, would change the speed of electrons which is not desirable in this experiment; because, it alters readers true understanding of the existence of spin separating force. So, the voltage difference used in the experiment must be a fixed voltage. High voltage of the experiment is provided by a flyback transformer. Such a transformer is usually used in video systems to generate the required high volatage and it consists of two parts: the first part includes a ferrite core and coil which boost the input alternating voltage to transformer (which depends on input frequency\footnote{\textit{Mistake in the old version (arXiv:1504.07963v1 and arXiv:1504.07963v2)}}); and the second part includes 6 diodes and some capacitors which convert the boosted alternating voltage to a constant DC voltage. In this experiment, no electric field and no magnetic field does not change with time. But if the electric field is changing with pulse moving electric charges will be quantized and Spin detachment seen better (Electrons have wave-like behavior) \\*
 \ignorespaces
\noindent\text{       } 
\textbf{So, as it is necessary to have a fixed voltage in this experiment, output of the flyback transformer must have a good quality.}
\noindent\text{       } 
At this experiment seems that there is no contradiction with principles of physics in this experiment, because the important factor is to observe the separation of electrons' speed which may result from Coulomb or other unknown forces that influence the path of electrons' flow and change their path from direct flow to spiral one. 
 \ignorespaces \noindent\text{       } 
At this experiment used a high vacuum lamp to observe the intrinsic spin of electrons (figure a-1). This lamp was made of three main parts: a high vacuum glass bulb (which is properly deionized), an electron gun and a phosphorescent sheet which includes high purity phosphorus. When the lamp was vacuumized, it is deionized using a diffusion pump. But you know that ions will not be totally removed from system; so atoms and molecules, even on a very small scale, are ionized and become highly reactive when electric field is applied. They react with phosphorus atoms which have been activated and make these atoms lose their initial purity and nature and the sheet is so called burned out. So deionization is almost a vital process and a certain type of alloy is used in higher levels to do deionization. This makes it possible to deionize the lamp through indirect heating. This alloy which is called Getter stands against the electron gun (figure 1-b, 1-c).\\*

\begin{center}
\noindent \includegraphics[width=150pt]{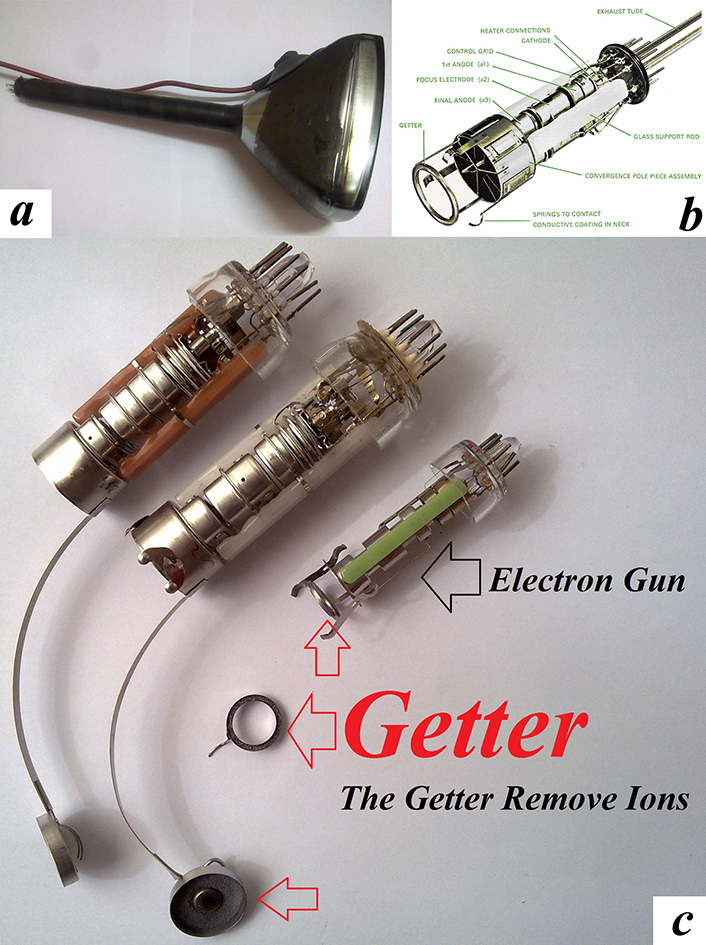}

\noindent\text{       } \ignorespaces
\noindent \textbf{Figure 1:} Figure a shows a used lamp; Figure b,c shows the structure of a electron gun and Getter.
\end{center}

\noindent\text{       } \ignorespaces
\noindent\text{       }  Besides, electrons and ions flow in opposite directions in the electric field and due to mass flow of electrons influenced by high voltage,you can ignore the presence of ions in the lamp. Electron beams, after being separated from filament, are paralleled by Grid Plates of gun which play the same role of collimators in Stern-Gerlach experiment, and then are shot to phosphorescent plate and hit it and cause illumination. The most sensitive part of the lamp is the electron gun and the related parts set much sensitively. Generally, this electron gun has some similarities and differences with Stern-Gerlach furnace. Both include some thermodynamic preparations for separation. When an electron beam is influenced by an appropriate potential difference, it will hit the screen and distributed in Guassian symmetric form. Of course, it is very important to set carefully the intensity by which beams hit the screen to obtain an appropriate Guassian form. When an electron beam is influenced by an inhomogeneous external magnetic field, spin elements are separated from each other. But two things are important here about electrons: firstly , as the mass of electron is small, Lorentz force cannot be easily removed. Secondly , obtaining a high magnetic field gradient will be a basic problem. Concerning the first problem, Lorentz force cannot be totally removed from the system but its effect on observation can be reduced to a great extent. In case of the second problem, there arises the important question that:\\*
\noindent\text{       }  \textbf{Is it necessary to have a very high magnetic field to produce a very high gradient?}  No. The value of magnetic field gradient is very high and indefinite near sharp pointed objects. Concerning this point, a simple geometric solution can be introduced. It is introduced based on the geometry used in magnets in the experiment conducted by Stern-Gerlach using Potassium atom beams
\cite{23},\cite{24}. Its truth was also proved in practice. Consider a circle as the center of all geometrical calculations (figure 2-a) to which enters a magnet with the radius a (1) and a second magnet is tangential to the circle's external plane (2). Caclulation of magnetic field lines is made using two points of intersection of the magnet (1) with circle calculator. Since high gradient is not required in Stern-Gerlach classical experiment using potassium atom beams, radius of the lower circle (magnet(1)) is a little smaller than the radius of main circle calculator.\\*
\noindent \textbf{Inhomogeneous magnetic field calculations:}\\*
Reference\cite{23} sec 3. Two-wire field   \\*\noindent\text{       } \ignorespaces
\noindent So long as the magnetization does not proceed to saturation, the pole pieces, of circular cylindrical form,lie in two equipotential surfaces of a two-wire system using currents in opposite directions. The magnetic field \textbf{\textit{H }}therefore consists of two components, \textit{H}${}_{1}$ and \textit{H}${}_{2}$ as shown in Figure 3:
\begin{equation}
H\left(r\right)=H\left(r_{1}\right)+H\left(r_{2}\right)
\label{eq.7}
\end{equation}
\begin{center}
\noindent \includegraphics[width=160pt]{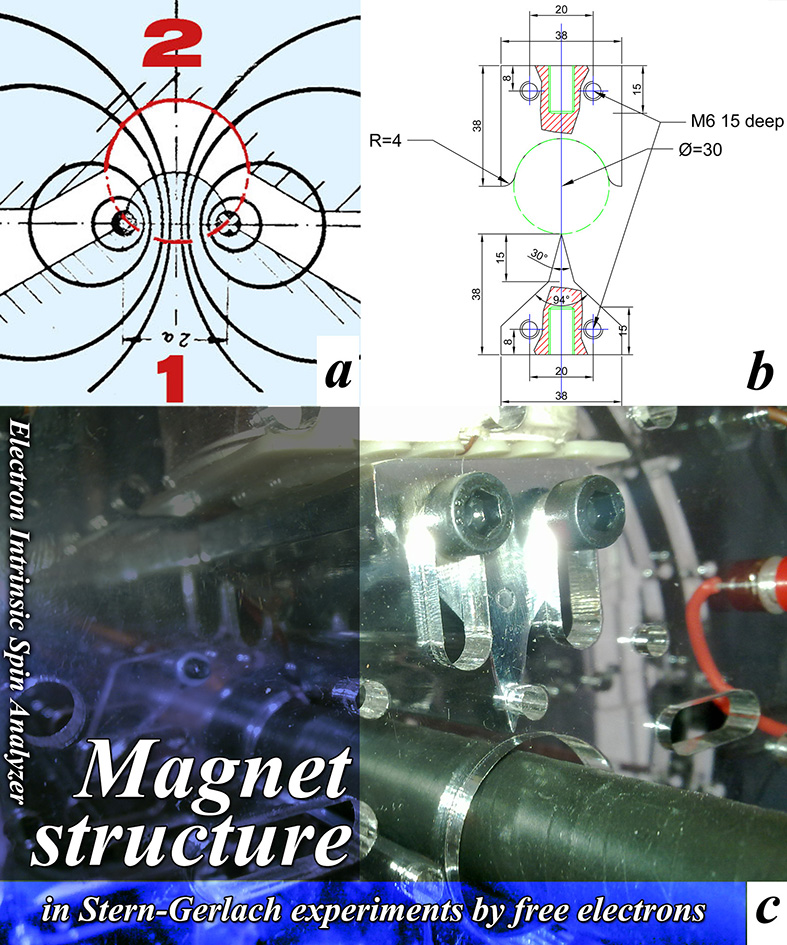}
\end{center}
\noindent\text{       } \ignorespaces
\noindent  \textbf{Figure 2:} Figure a represents the magnet used in Stern-Gerlach classical experiment on potassium atom beams\cite{23},\cite{24},Figure b-c represents the magnet used in classical Stern-Gerlach experiment on free electron beams. \\*
\begin{center}
\noindent \includegraphics*[width=170pt]{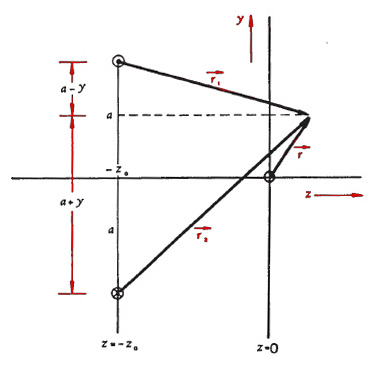}
\end{center}
\noindent\text{       } \ignorespaces
\noindent \textbf{Figure 3:} Determination of a system of coordinates\\*

\noindent\text{       } \ignorespaces
\noindent Each of the two conductors contributes to the field as follows:
\begin{equation}
H_{i}\left(r_{i}\right)=\frac{I_{i}\times r_{i}}{2\pi r_{i}^2}(i=1,2)
\label{eq.8}
\end{equation}
Where 
\begin{equation}
I_{1}=-I_{2}=I
\label{eq.9}
\end{equation}
is the excitation current for the magnetic fields. Hence, at the point \textit{r },
\begin{equation}
H\left(r\right)=\frac{2}{2\pi }I\times \left(\frac{r_{1}}{r_{1}^2}-\frac{r_{2}}{r_{2}^2}\right)
\label{eq.10}
\end{equation}
The value of the magnetic field strength is obtained by squaring this expression. remembering in the subsequent calculation that \textit{r}${}_{1}$ and \textit{r}${}_{2}$ lie in a plane at right angles to \textit{l, }one finally obtains:
\begin{equation}
H=\frac{I}{\pi }\frac{a}{r_{1}r_{2}}
\label{eq.11}
\end{equation}
The change in the value of H as a function of z can be calculated, using
\begin{equation}
r_{1}^2={\left(a-y\right)}^2+{\left(z+z_{0}\right)}^2
\label{eq.12}
\end{equation}
And
\begin{equation}
r_{2}^2={\left(a+y\right)}^2+{\left(z+z_{0}\right)}^2
\label{eq.13}
\end{equation}
as
\begin{align}
\frac{\partial H}{\partial z}&=-\frac{I.a{\left(z+z_{0}\right)}^2}{\pi }\times\frac{\left(r_{1}^2+r_{2}^2\right)}{r_{1}^3r_{2}^3}=\nonumber \\
&=\frac{2I.a{\left(z+z_{0}\right)}^2}{\pi }\times\nonumber \\
&\times\frac{a^2+y^2+{\left(z+z_{0}\right)}^2}{{\left({\left(a^2-y^2\right)}^2+2{\left(z+z_{0}\right)}^2 \left(a^2+y^2\right)+{\left(z+z_{0}\right)}^4\right)}^{\frac{3}{2}}}\nonumber \\
\label{eq.14}
\end{align}

The surfaces of constant field inhomogeneity are shown in (Figure 4).
\begin{center}
\noindent \includegraphics[width=170pt]{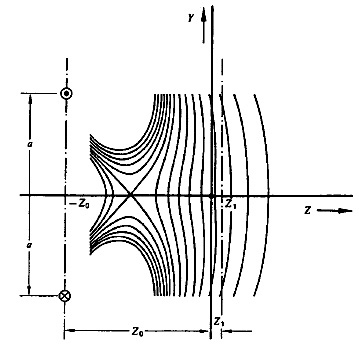}\\*
\noindent\text{       } \ignorespaces
\noindent \textbf{Figure 4:}  Lines of constant field inhomogeneity.\\*
\end{center}

\noindent\text{       } \ignorespaces
\noindent The equipotential surfaces in the neighbourhood of  \textit{z = z${}_{1}$, }are to be regarded as planes, to a good approximation. it must now find the plane \textit{z = z${}_{1}$}, in which the equipotential surfaces are as plane as possible, and how far this plane lies from the plane containing the wires, \textit{z}=\textit{z}${}_{0 }$.To do this, the length of the element of path (\textit{z}${}_{0}$+ \textit{z}${}_{1}$) will be determined, subject to the condition that,in the neighbourhood of \textit{y}=0, $\frac{\partial H}{\partial z}$\textit{ }is independent of \textit{y.} If one develops $\frac{\partial H}{\partial z}$\textit{ }in a series of y${}_{2}$ and breaks this off after the first order, on the assumption that y${}_{2}$ is small compared with (\textit{z}${}_{0}$+ \textit{z}${}_{1}$)${}^{2}$ or a${}_{2}$, the field gradients are found to be
\noindent \textit{}
\begin{equation}
\left|\frac{\partial H}{\partial z}\right|_{z_{1}}=\frac{2I.a\left(z_{0}+z_{1}\right)}{\pi }
\label{eq.15}
\end{equation}
Dependence on y is to vanish at \textit{z = z${}_{1}$}. We then get: 
\begin{equation}
2a^2-{\left(z_{0}+z_{1}\right)}^2=0
\label{eq.16}
\end{equation}
from which is follows that
\begin{equation}
z_{1}+z_{2}=a.\sqrt{2}
\label{eq.17}
\end{equation}
\noindent\text{       } \ignorespaces
The field inhomogeneity begins to decrease steeply with increasing y only at greater distances along the z-axis. The present apparatus has a diaphragm system in which the length of the radiation window is about 4/3 a. 
\begin{center}
\noindent \includegraphics*[width=140pt]{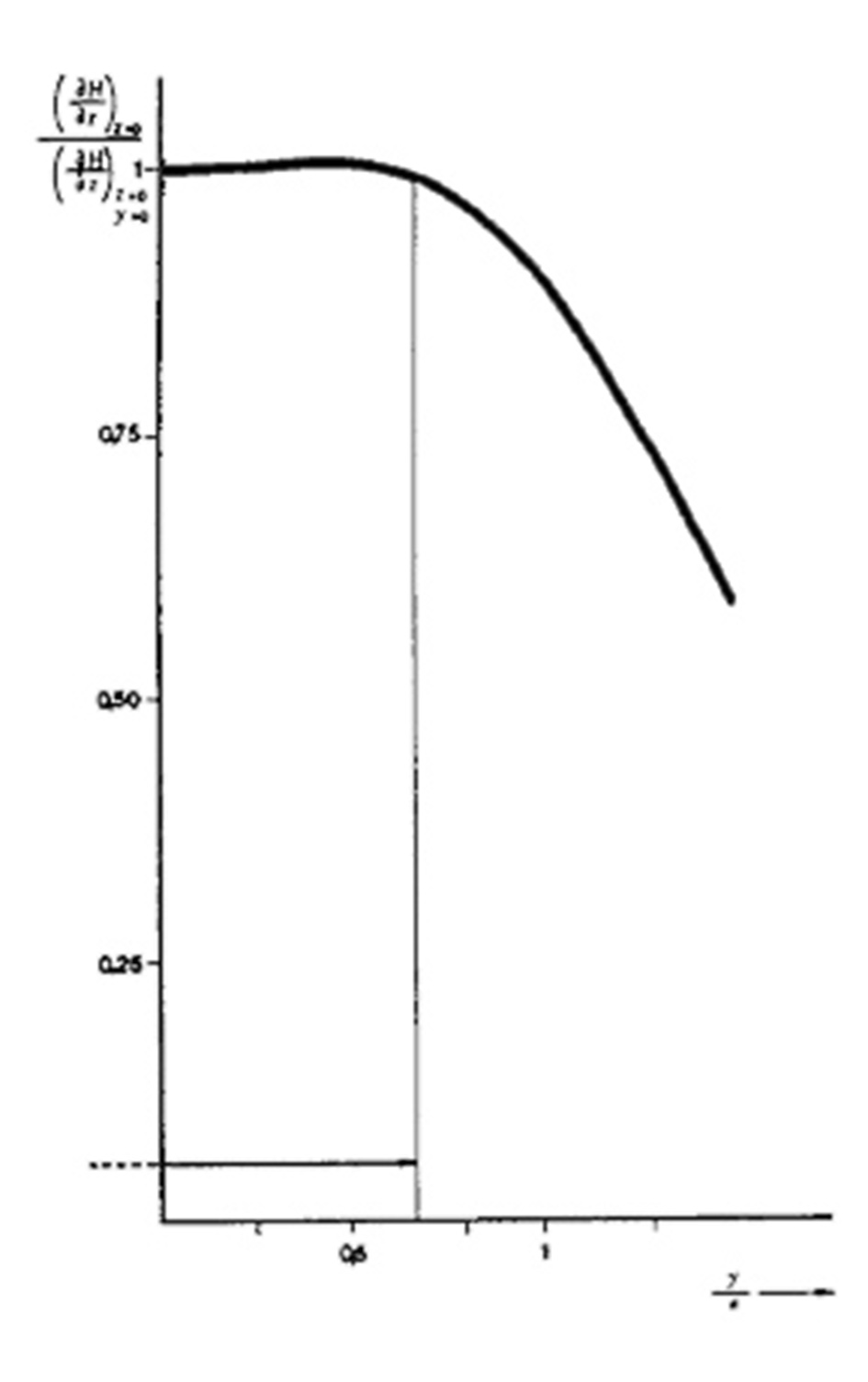}
\end{center}
\noindent \textbf{Figure 5:} Behaviour of field inhomogeneity along the radiation window.\\*
\\*
\noindent\text{       } \ignorespaces
\noindent As Fig 5 shows, the value of $\frac{\partial H}{\partial z}\ $at \textit{y}=$\frac{2}{3}$ \textit{a }scarcely differs from its value at y=0. The condition for constant inhomogeneity is thus met to a large extent. Now only \textit{H }can be measured in the region of the \textit{z}-axis, and not $\frac{\partial H}{\partial z}$\textit{ }. Hence it is also useful to find that plane for which

\begin{equation}
\left|\frac{\partial H}{\partial z}\right|=\frac{a}{H}=\varepsilon 
\label{eq.18}
\end{equation}
\noindent\text{       } \ignorespaces
is a value not depending on y in the neighbourhood of  y = 0. This plane is to be z = 0, i. e., it helps to fix z${}_{0}$. Expansion as a series in y${}^{2 }$gives

\begin{align}
\varepsilon &=\frac{2a\left(z+z_{0}\right)}{a^2+{(z+z_{0})}^2}\times \nonumber \\
& \times\left(1+\frac{y^2}{{\left(a^2+{\left(z+z_{0}\right)}^2\right)}^2}.\left(5a^2-3{\left(z+z_{0}\right)}^2\right)\right)
\label{eq.19}
\end{align}
Dependence on y should vanish at z = 0. We then get

\begin{equation}
4a^2-3z_{0}^2=0
\label{eq.20}
\end{equation}
 from which it follows that 

\begin{equation}
z_{0}=a\sqrt{\frac{5}{3}}=1.29\ a
\label{eq.21}
\end{equation} 
Hence
\begin{equation}
z_{1}=\left(\sqrt{2}-\sqrt{\frac{5}{3}}\right)a=0.12a\ll z_{0}
\label{eq.22}
\end{equation}
\noindent\text{       } \ignorespaces
The plane \textit{z = z${}_{1}$ }lies therefore immediately adjacent to the plane \textit{z = 0}. Hence the inhomogeneity at \textit{z = 0 }can be regarded as being constant, to a good approximation. The Stern-Gerlach apparatus is adjusted, in view of the foregoing relationships, so that the radiation window lies around 1.3 \textit{a }from the notional wires of the two-wire system (Fig 4).The calibration \textit{H(i) }of the electromagnet (magnetic field \textit{H }of magnetic induction \textit{B }against the excitation current \textit{i }) is likewise assumed for \textit{z }=1.3 \textit{a }.  The constant \textit{$\varepsilon $ }can therefore be calculated from
\begin{equation}
\varepsilon \left(z=0\right)=\frac{2.\sqrt{\frac{5}{3}}}{1+\frac{5}{3}}=0.968
\label{eq.23}
\end{equation}
\noindent\text{       } \ignorespaces
Field strengths are therefore converted to field gradients using the equation
\begin{equation}
\left|\frac{\partial H}{\partial z}\right|=0.968\frac{H}{a}
\label{eq.24}
\end{equation}
\noindent\text{       } \ignorespaces
 According to the above mentioned in\cite{24}, magnetic field gradient will finally be calculated in the following form (for the center of circle calculator) \\*
\begin{equation}
\frac{\partial H}{\partial z}=0.968\frac{H}{a}
\label{eq.25}
\end{equation}
and the simplified form will be as: \\*
\begin{equation}
\frac{\partial B}{\partial z}=\frac{B}{a}
\label{eq.26}
\end{equation}
\noindent\text{ }  In this equation \eqref{eq.26}, final magnetic field gradient has an inverse relationship with the radius of internal circle and yet a direct relationship with the value of magnetic field applied on magnets. Although this equation seems simple, it can help us understand how much magnetic field gradient varies with the variation of internal magnet radius (1). Now suppose that the internal magnet radius (1) goes to zero (a limited to zero), then the radius of circle will be very low in equation an in fact you will have a very sharp magnet, the small number in denominator will turn to a large coefficient in the field size, and the final inhomogeneous magnetic field gradient will significsntly increase. Actually, such magnets were designed (Figure 2-c, 2-b) and one of them was sharp pointed so that when located on a certain point separation of spins could be observed. It is noteworthy that the direction of magnetic field lines is an important factor for recognizing this phenomenon. The sharp point of one magnet must be N and the curve end of another one must be S.\\*
\noindent\text{       } \ignorespaces
  Actually, by applying inhomogeneous magnetic field, electron beams are deviated to Lorentz force direction, spin force will separate electrons' spin and split the initial beam into two beams which can be easily observed by eyes and are at close distance to each other. By increasing the strength of magnetic field, beams will get more away compared to their initial distance and have a better image of separation. The image quality of spin separation depends on three factors: electric field, and the distance of magnets from exit aperture of electron gun. \\*
\noindent\text{       } \ignorespaces
A question may come to your mind that:\\*{}
\noindent\text{       } \ignorespaces
\textbf{Can objective observation be the result of Lorentz force?} Since the mass of electrons and the electric field is very small, Lorentz force cannot be removed from system but it can be observed simultaneously in the experiment and distinguished according to the extent which electron beams travel. What can be observed in practice is that when inhomogeneous magnetic field is applied, the initial single electron beam split into two beams very close to each other (10 micro meter to 100 micro meter) and the two beams travel together(at the one time). Besides, due to size of the magnetic field used in the experiment, Lorentz force exerted a deviation of some millimeters or about one or some centimeters, the extent of which can be easily distinguished from separating force of electron's spin.\\*{}
\noindent\text{       } \ignorespaces
\textbf{Beams are split into two beams only when the applied magnetic field is inhomogeneous}.(figure 6-a, 6-b).\\*
\begin{center}
\includegraphics[width=225pt]{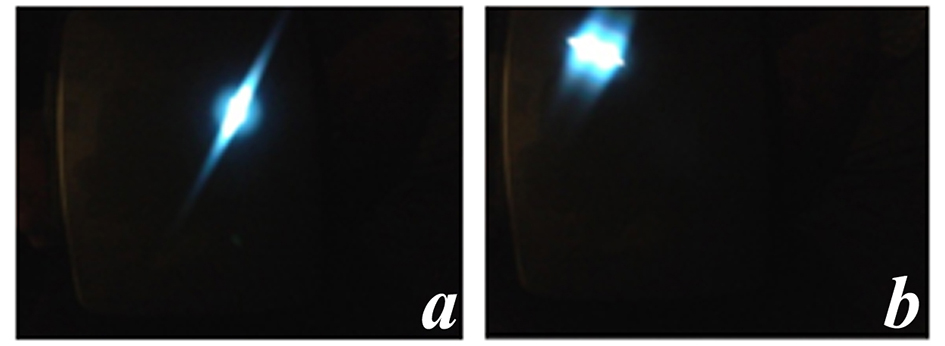}
\end{center}
\noindent\text{       } \ignorespaces
\noindent \textbf{Figure 6:} Figure a represents an electron beam before applying an inhomogeneous magnetic field. Figure b represents a pair-beam electron after applying the inhomogeneous magnetic field (using a stationary magnet) and the two beams travel together. \\*
\\*
\noindent\text{   }\ignorespaces \textbf{May the objective observation result from variation of electron velocity distribution?(different velocity)}  It's important note. If there is variation in electron velocity distribution, it may seem as separation(?).\\*
\noindent\text{       } \ignorespaces
 But this may never happen in this experiment as the electric field is constant (according to the electron velocity order). to This can be tested in practice and convinced that if two flat magnets are used instead of gradient magnets, you can expect that electrons travel exactly according to Lorentz force and if there is variation in electron velocity distribution, it will be proved well. 
This hypothesis was tested and found that when flat magnets are used: the single-beams merely travel according to Lorentz force and no split was reported. So this would occur only in the presence of gradient and all electrons have the same speed.\\*
\noindent\text{       } \ignorespaces \textbf{\large So,} Beams are split into two beams only when the applied magnetic field is inhomogeneous.\\*
\noindent\text{       } \ignorespaces \textbf{Does the direction of electric field lines change?} No, electric field voltage is fed and strengthened by diodes and capacitors.\\*
\noindent\text{  } \noindent you observed that along the widthwise direction of electrons emission, too much static electricity is generated which is a good evidence that suggests it is very likely that electrons do not travel in a quite direct path. Besides, this can clearly be observed when voltage is reduced to a certain extent. The charge of static electricity around the lamp greatly affect the path of electrons so that it requires to place objects close to the earth in order to eliminate this effect and avoid any intervention in the result. As voltage increases, the quality of paired image tends to decrease so that there will no longer be enough resolution. When the voltage is very low, no image can be observed. It is evident that as electric field changes, the length of electron-bunch train emitted from gun will change and this in turn affect the quality and resolution of the image. By applying an inhomogeneous magnetic field, distribution of Guassian beam changes. So far, all calculations were made based on the hypothesis that electrons travel through a direct path in the space which originates from a classical view. On the other hand, all calculations made on electron velocity so far focused on single electrons, while there is no single or individual electron in reality. So one consequence for holding this view would be to ignore coulomb force or other similar forces. Therefore, when some electrons travel together closely, the repulsion and attraction between them place them in a balanced distance to each other. If an electric field, under a certain voltage, make electrons move along the spiral pattern in space (figure 7-a) one can claim that electron impulses before applying the inhomogeneous magnetic field would not be merely limited to moving along the direction of electrons. But components of the impulse axis for each electron is polarized along with the screen for electrons flow. So the impulses will be significantly reduced along with the electron emission. Hence, it is evident that electrons will have much more time to interact with inhomogeneous magnetic field (figure 7-b) and you can see more separation distance which was seen in practice. \\*
\begin{center}
\noindent \includegraphics[width=170pt]{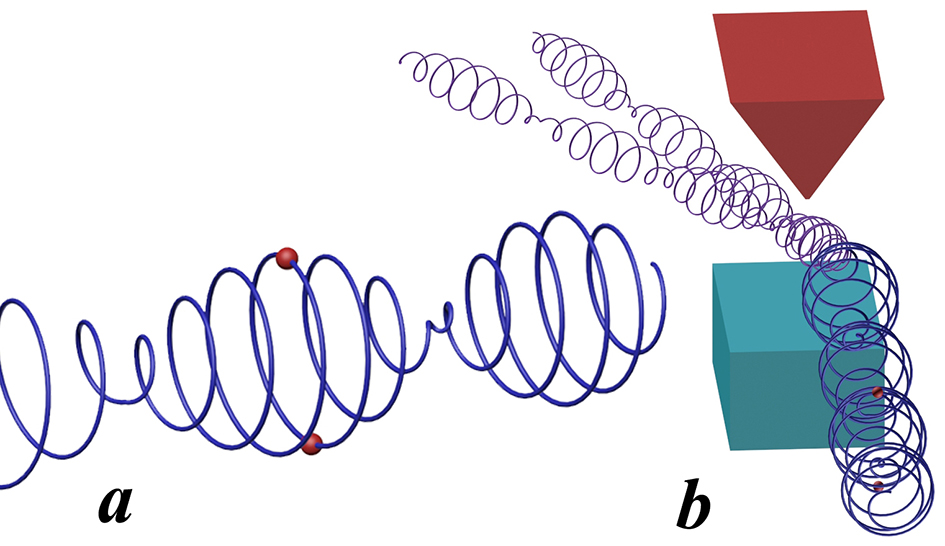}
\end{center}
\noindent\text{       } \ignorespaces
\noindent \textbf{Figure 7:}  Figure 7-a represents the spiral path of electrons caused by interaction. Figure b shows  the spiral path of electrons when confronted with inhomogeneous magnetic field.\\*
\end{multicols}
\begin{center}
\begin{tabular}{p{1in}p{1in}p{1in}p{1.0in}p{1in}p{0.7in}} \hline 
\textbf\tiny{\small Accelerating \newline Voltage\newline  (in KV){}} 
& \textbf\tiny{\small Velocity\newline of electerons \newline (m/sec){}}
& \textbf\tiny{\small Mass\newline of Electerons\newline  (Kg){}}
& \textbf\tiny{\small Energy\newline of Electerons\newline (N*m){}} 
& \textbf\tiny{\small Momentum\newline of Electrons\newline (N*sec){}}
& \textbf\tiny{\small Wavelength \newline (m){}} \\ \hline 
\small 5{} &\small 4.19E+07{} &\small 9.11E-31{} &\small{} 8.01E-16 &\small 3.82E-23 &\small 1.73E-11 \\
\small 6 {}&\small 4.59E+07 &\small 9.11E-31 &\small 9.61E-16 &\small 4.18E-23 &\small 1.58E-11 \\ 
\small 7{} &\small 4.96E+07 &\small 9.11E-31 &\small 1.12E-15 &\small 4.52E-23 &\small 1.47E-11 \\ 
\small 8{} &\small 5.30E+07 &\small 9.11E-31 &\small 1.28E-15 &\small 4.83E-23 &\small 1.37E-11 \\ 
\small 9{} &\small 5.63E+07 &\small 9.11E-31 &\small 1.44E-15 &\small 5.13E-23 &\small 1.29E-11 \\ 
\small 10{} &\small 5.93E+07 &\small 9.11E-31 &\small 1.60E-15 &\small 5.40E-23 &\small 1.23E-11 \\ 
\small 12{} &\small 6.50E+07 &\small 9.11E-31 &\small 1.92E-15 &\small 5.92E-23 &\small 1.12E-11 \\ 
\small 13{} &\small 6.76E+07 &\small 9.11E-31 &\small 2.08E-15 &\small 6.16E-23 &\small 1.08E-11 \\  
\small 14{} &\small 7.02E+07 &\small 9.11E-31 &\small 2.24E-15 &\small 6.39E-23 &\small 1.04E-11 \\ 
\small 15{} &\small 7.26E+07 &\small 9.11E-31 &\small 2.40E-15 &\small 6.62E-23 &\small 1.00E-11 \\ 
\small 16 {}&\small 7.50E+07 &\small 9.11E-31 &\small 2.56E-15 &\small 6.83E-23 &\small 9.70E-12 \\ 
\small 17 {}&\small 7.73E+07 &\small 9.11E-31 &\small 2.72E-15 &\small 7.04E-23 &\small 9.41E-12 \\  
\small 18 {}& \small7.96E+07 &\small 9.11E-31 &\small 2.88E-15 &\small 7.25E-23 &\small 9.14E-12 \\  
\small 19 {}&\small 8.17E+07 &\small 9.11E-31 &\small 3.04E-15 &\small 7.45E-23 &\small 8.90E-12 \\ 
\small 20{} & \small8.39E+07 &\small 9.11E-31 &\small 3.20E-15 &\small 7.64E-23 &\small 8.67E-12 \\ 
\small 21{} &\small 8.59E+07 &\small 9.11E-31 &\small 3.36E-15 &\small 7.83E-23 &\small 8.46E-12 \\ 
\small 22{} & \small8.80E+07 &\small 9.11E-31 &\small 3.52E-15 &\small 8.01E-23 &\small 8.27E-12 \\  
\small 23{} &\small 8.99E+07 &\small 9.11E-31 &\small 3.68E-15 &\small 8.19E-23 &\small 8.09E-12 \\  
\small 24{}&\small 9.19E+07 &\small 9.11E-31 &\small 3.84E-15 &\small 8.37E-23 &\small 7.92E-12 \\ 
\small 25 {}&\small 9.38E+07 &\small 9.11E-31 &\small 4.01E-15 &\small 8.54E-23 &\small 7.76E-12 \\ 
\small 26{} &\small 9.56E+07 &\small 9.11E-31 &\small 4.17E-15 &\small 8.71E-23 &\small 7.61E-12 \\ 
\small 27{} &\small 9.75E+07 &\small 9.11E-31 &\small 4.33E-15 &\small 8.88E-23 &\small 7.46E-12 \\ 
\small 28{} &\small 9.92E+07 &\small 9.11E-31 &\small 4.49E-15 &\small 9.04E-23 &\small 7.33E-12 \\ 
\small 29{} &\small 1.01E+08 &\small 9.11E-31 &\small 4.65E-15 &\small 9.20E-23 &\small 7.20E-12 \\ 
\small 30{} &\small 1.03E+08 &\small 9.11E-31 &\small 4.81E-15 &\small 9.36E-23 &\small 7.08E-12 \\ \hline 
\end{tabular}\\*
\begin{center}
\text{}
\\*
\textbf{Table 1:} shows the values of voltage, velocity, mass, energy, impulse and wavelength.
\end{center}
\end{center}
\noindent 
\ignorespaces
\begin{multicols}{2}
\noindent\text{       } \ignorespaces So what you know as drift velocity of electrons and is related to ($\approx10^5 m/s$) can refer to the speed along with electrons emission not speed in its cross section. Of course more research is required to measure the actual speed in the cross section. Images obtained in this experiment conform a lot to the experiment simulated in some articles which used the wave packet concept.\cite{18},\cite{22} Using a CCD camera and a similar system and do the fitting by matlab or some other professional software, you can prove its objective truth in practice. Voltage of the lamp used in this experiment ranged from 5kV to 25kV . Electrons velocity, energy range, and the related wave length have been discussed in table 1. unfortunately, due to the measuring instrument it reported voltage span.\cite{21}\\*
 \\*
\noindent Concerning the value of velocity which is from 
($\approx10^7 m/s$) ,so $\gamma\approx 1$
\begin{equation}
\gamma{\rm =}\frac{1}{\sqrt{{\rm 1}{\rm -}{{\rm (}\frac{{\rm v}}{{\rm c}}{\rm )}}^{{\rm 2}}}}{\rm =}\frac{1}{\sqrt{{\rm 1}{\rm -}{{\rm (}\frac{?.?{\rm \ \sim }{{\rm 10}}^{{\rm 7}}}{{\rm c}}{\rm )}}^{{\rm 2}}}}\approx 1
\label{eq.27}
\end{equation}
\noindent\text{       } \ignorespaces
so velocity cannot be proportional. The electric field will be ($\approx10^5 V/m$) separation distance of electron beam is about($\approx10 \mu m$) to ($\approx100 \mu m$). This separation can be observed in the following figures.
(figure 8)
\begin{center}
\noindent \includegraphics[width=190pt]{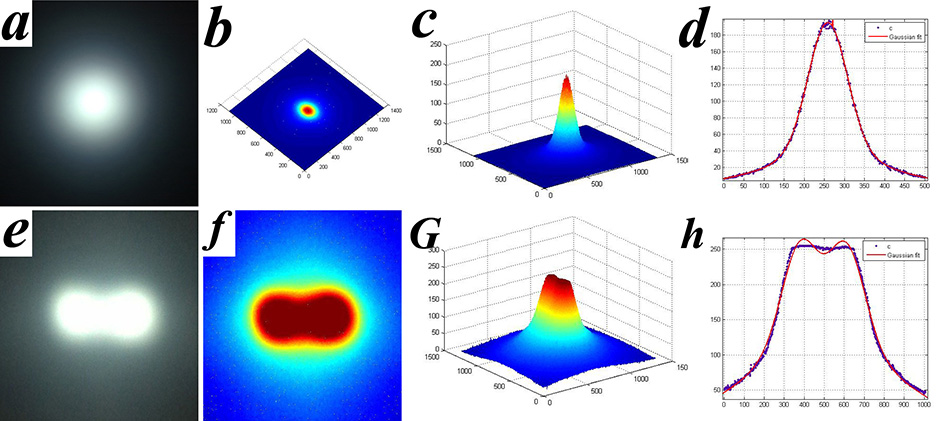}
\end{center}
\noindent\text{       } \ignorespaces
\noindent \textbf{Figure 8:}Figure a,b,c,d represent the electron beam before applying inhomogeneous magnetic field. e,f,g,h represent the electron beam after applying inhomogeneous magnetic field recognized by DC current.\\
\noindent\text{       }\noindent Spin separation can be better observed using a linearizer which creates a secondary inhomogeneous magnetic field (figure 9).\\*
\begin{center}
\noindent \includegraphics[width=190pt]{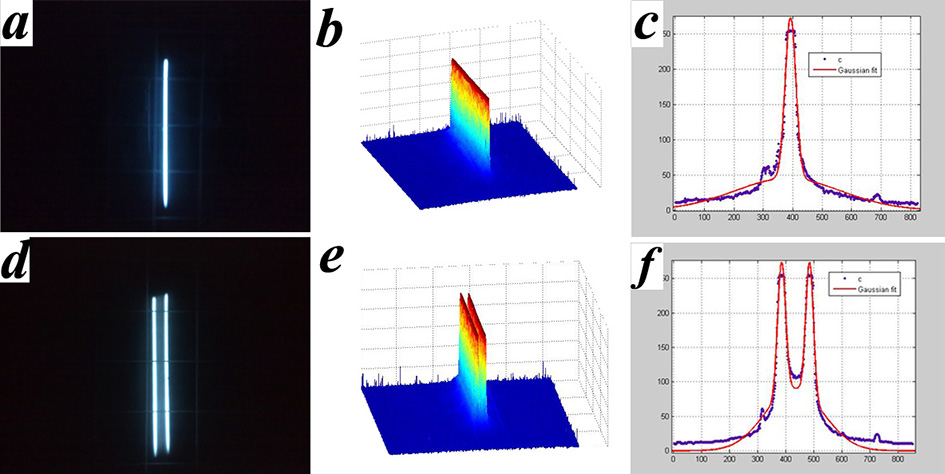}
\end{center}
\noindent\text{       } \ignorespaces
\noindent \textbf{Figure 9:} Figure a,b,c represent linearized electron beam before applying magnetic field. Figure d,e,f  represent the electron beam which has turned to parallel lines after applying magnetic field.\\*
\noindent Figure 10 displays a general view of the machine.
\begin{center}
\noindent \includegraphics[width=190pt]{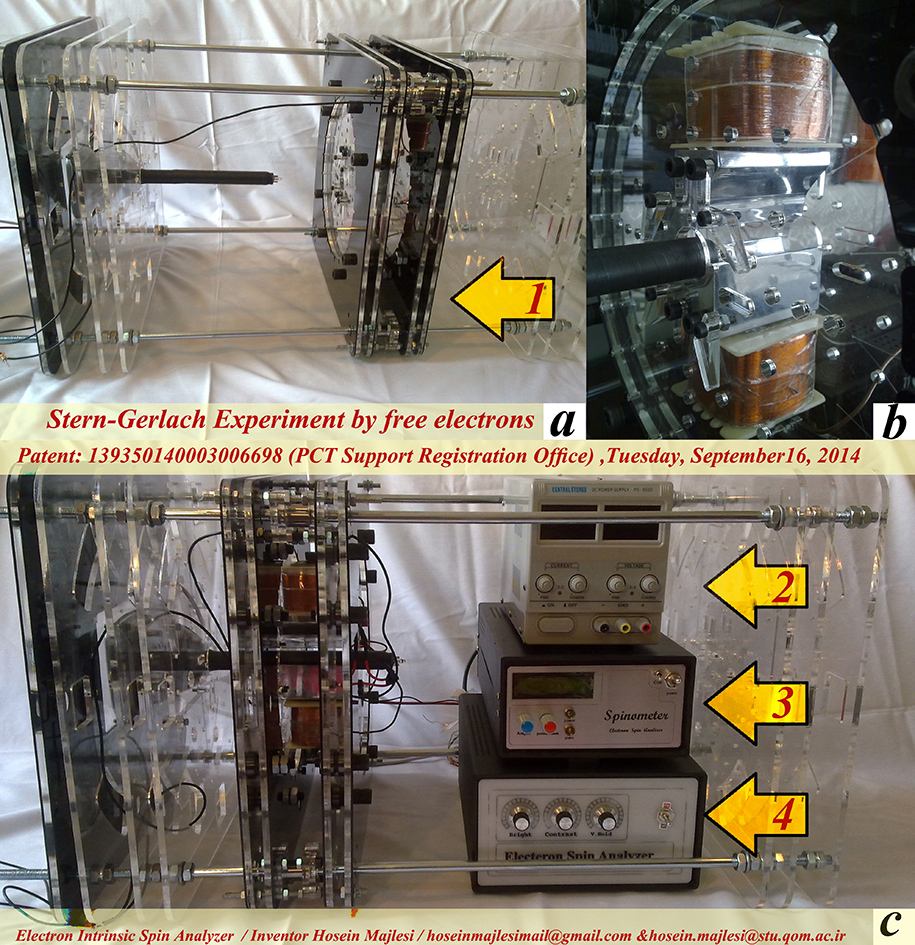}
\end{center}
\noindent\text{       } \ignorespaces
\noindent \textbf{Figure 10:} Figure a displays a general view of magnets and the chassis where magnets can easily move back and forth and their distance from electron gun can be adjusted. Of course, more consecutive magnets can be used in the experiment. (successive Stern-Gerlach experiment), Figure b displays a view of magnets while they can easily rotate around the pipe and be adjusted and fixed in a certain. Figure c displays a general view of the machine and all the related parts. Flash 1 shows the moving frame of magnets. Flash 2 shows the feeding source of DC. This source provides us with a constant and straight DC current. Of course, you can replace it with a stationary magnet to produce magnet field gradient.  Flashes 3 and 4 represent the feeding source of electric field of a lamp.\\*
\noindent  \\
\noindent \textbf{\large\bf Results}\\
Stern-Gerlach experiment using the electron beam had not already been performed and contrary to what Bohr and Pauli predicted, observation of electrons spin was not only possible but it also happened in practice and it conformed to classical equations as well. Spin separation occur within a certain voltage span so that if the electrons' voltage is greater than a certain amount, the phenomenon cannot be observed due to the short time interaction between electrons and magnetic field and the image of separation does not have much resolution. If voltage is lower than a certain amount, beams will never approach the screen and there will be no image on the screen. To product a high magnetic field gradient does not necessarily require a strong magnetic field and it can be created within a certain distance from a sharp pointed magnet in a certain form. Besides, it was not enough to have only a high magnetic field gradient. It was also important to use an appropriate voltage. The quality of spin separation image improves by adding three components and beams will be more distinguished: Electric field and the strength of magnetic field (degree of gradient of magnetic field) and increasing the distance of magnets from the exit aperture of electron gun, ... Contrary to what was expected before about certain voltage span, electrons do not travel in a direct path in the space due to coulomb force or other unknown forces. They are also emitted in the cross section and emit lots of static electricity charge around. it was very similar to simulations\cite{18} ,\cite{22} Electron beams can be observed with a better quality after separation using a linearizer. that objective observation requires your consideration of some technical points simultaneously.  \\*
\noindent \textbf{\large\bf Acknowledgement }\\
\noindent\text{  }\ignorespaces
I thank Khajeh Nasir Toosi University that pursued the invention on the side of Patent Registration Office. 
I am also thankful to 
Ms.Mohamadi who carried out the translation..

\noindent \textbf{Figure References: } Figure 1-b:Internet Figure 2-a,3,4,5 :\cite{23}\\*
\end{multicols}
 \end{document}